\documentclass[aps,twocolumn,prd]{revtex4-1}
\pdfoutput=1

\usepackage{color}
\usepackage{slashed}
\usepackage{hyperref}
\usepackage{amsmath}
\usepackage{amssymb}
\usepackage{latexsym}
\usepackage{amsfonts}
\usepackage{epsfig}
\usepackage{psfrag}
\usepackage{graphicx}
\usepackage{mathtools}

\def\hhref#1{\href{http://arxiv.org/abs/#1}{arXiv:#1}} 

\newcommand{\bea}{\begin{eqnarray}}
\newcommand{\ea}{\end{eqnarray}}
\newcommand{\eea}{\end{eqnarray}}

\begin{document}

\title{Vacuum decay and the transmission resonances in space-dependent electric fields}
\author{Cesim~K.~Dumlu}
\affiliation{Department of Physics, Middle East Technical University, 06800, Ankara, Turkey}
\email{cdumlu@metu.edu.tr}

\begin{abstract}
We investigate the decay of quantum electrodynamical (QED) vacuum  in arbitrary space-dependent electric fields. In particular, we analyze the resonance peaks of the positron emission spectrum for the external fields with subcycle structure. For this, we study the transmission probability in the framework of the scattering approach to vacuum pair production. In under-the-barrier scattering regime, we show that the width of a transmission resonance can be enhanced when the effective scattering potential contains multiple wells. Such a broadening in the resonance width corresponds to a decrease in the tunneling time. This may be relevant for observing the vacuum decay at shorter time scales before the external field is adiabatically turned off. In above-the-barrier scattering regime, we give a set of coupled differential equations for the numerical computation of the Bogoliubov coefficients.
\end{abstract}

\date{\today}

\pacs{
03.65.Sq, 	
12.20.Ds, 
11.15.Tk, 
}

\maketitle

\section{Introduction}

The decay of vacuum into electron-positron pairs under the influence of an external electric field is a remarkable prediction of QED \cite{sauter,schw,he}. In a constant electric field $E_0$, Schwinger found the vacuum decay rate per unit volume per unit time as \cite{schw}:($\hbar=c=1$)
\begin{eqnarray}
\mathcal{P}=
\frac{q^2 E_0^2}{4 \pi^3}\sum^{\infty}_{n=1}\frac{1}{n^2}\, \text{exp}\left[-\frac{n\pi m^2}{qE_0}\right],
\label{schw}
\end{eqnarray}
where $m$ and $q$ are the electron's mass and charge respectively. The threshold intensity for the electric field to create an appreciable amount of pairs is $E_{\text{cr}}\sim m^2/q =10^{18}\text{V/m}$.  The experimental verification has been long coming because of such  a high threshold. In the near future, it is hoped that advances in strong laser pulses will bring us close to intensities as great as $10^{-3}E_{\text{cr}}\,$\cite{eli}. This value is however, still far from the observational regime due to the nonperturbative nature of the phenomenon: the exponent in (\ref{schw}) contains the inverse of power of $E_0$.This makes the decay rate in such a background strongly suppressed. Consequently, there have been investigations on lowering the pair production threshold by a combination of multiple pulses with varying time scales, intensities and polarizations \cite{dunne1,dunne2,nazo1,gies1,pizza1,bellnazo,anton}. These efforts accumulated valuable results. For instance, the focusing of multiple Gaussian beams to a single focal point  gives rise to a significant reduction in the threshold energy \cite{nazo1}. The subcycle structure of the time dependent electric fields plays an important role in the context of prolific pair production.  It has been observed that pair creation rate can be enhanced when a fast varying weak pulse is superimposed with a slowly varying strong  pulse \cite{dunne2}. The investigation of time-alternating fields with additional parameters such as the carrier phase and the chirp revealed that strong interference effects may occur for the certain modes of the created pairs \cite{gies1,flor1,dumlu3}.

On the other hand, the discussion of vacuum decay in spatially inhomogeneous fields was mainly considered for the cases, where the external field is represented by a smooth steplike barrier \cite{niki1,niki2,hansen,gies2,kimpage,remo}, or by a single potential well \cite{greinerb,dombey}.  In the latter case transmission resonances occur in the positron emission spectrum \cite{greinerb}. Such resonances widely appear in the scattering problems when the external potential supports quasi-bound states with complex energies \cite{mois}. These states are metastable; they decay in time due to the imaginary part of their energy. A metastable state in an external electric field may be thought of as  a quasi-particle excitation of the vacuum with complex energy $\epsilon=\omega -i \Gamma$. In general, the real part of the energy gives the position of the resonance peak in the spectrum and, it matches with the bound state energy level of the well in the effective scattering potential.  The imaginary part of the energy gives the resonance width $\Gamma \sim \Delta\,\omega$, and sets a natural timescale for the problem. This timescale is given by $t_t \sim 1/\Gamma$ and corresponds to tunneling escape rate of the trapped states from the well \cite{sakaki}. In the time limit $t > t_t$, when a metastable state decays into continuum, vacuum decay rate is identified with the transmission probability.

In the case where potential well supports multiple quasi-bound states, one might be tempted to think that $t_t$ is dominated by the state with the smallest imaginary energy. On the other hand, it was recently shown that $t_t$ is in fact characterized by the collective contribution of each quasi-bound state such that $t_t \approx 1/\sum_{n} \Gamma_n$ \cite{grob}. One important consequence of this is that the positron spectrum with larger resonance peaks may be resolved at relatively shorter time scales.
In conjunction with this, we show that the width of the resonance peaks in the transmission probability can be significantly enhanced if the effective scattering potential contains multiple wells.  Such an enhancement in the resonance width could be relevant for the observation of vacuum decay at shorter time scales with lower intensities and therefore, subcycle structure of the electric field must be taken into consideration. This however presents difficulties as far as the practical applications are concerned. One main restriction is that in order to talk about the tunneling time, one must either  assume the intensity of the external field remains constant over a timescale which is larger than $t_t$, or the variation in the intensity with respect to time is negligible. On the other hand, in realistic pulse configurations the scale of
spatial and temporal variation are of the same order, which is set by
the laser frequency. To be able to work with a time-independent electric field, we consider a static source such as a charge distribution $\rho(\vec{r})$ in equilibrium. A real example of such a system would be an ionic crystal with charged layers. For instance, in ferroelectric crystalline structures  an alternating electric field exists in the plane perpendicular to the atomic layers, where the electrostatic potential satisfies periodic boundary conditions \cite{meyer}. The charge density on the layers could be as high as to give an electric field intensity of $\sim 10^{-7} E_{\text{cr}}$. However in those systems the dispersion relation for quasi-particle excitations is generally nonlinear in momentum, so it remains doubtful whether such crystal structures can be modeled as the QED vacuum for Dirac particles as in the case of graphene \cite{schg}.

In the remainder of this work we will assume an idealized scenario where background field is purely
space-dependent. In the first part of the Sec. II, we work out the transmission probability in an arbitrary external potential in the framework of the scattering approach to vacuum decay. We use the semiclassical approximation to find a closed form expression for the transmission probability, when the energy is below the barrier. Our result is general and simple to use, and may be applied to the other areas such as graphene based superlattices, where transmission resonances are important \cite{gra1,gra2,gra3}.   In the remainder of Section II, we compare the transmission spectrum for the single and triple well configurations and show that in the latter case specific resonance widths can be broadened by adjusting the width of the wells. Following this we present the equations  which are suitable for numerical integration of the Bogoliubov coefficients, when the energy is above the barrier. We compare the analytical and the numerical results for an exactly solvable case. The final section contains our comments and conclusions. In an Appendix, we lay out the rules for obtaining the general form of the transmission probability and give the explicit results for a symmetric scattering potential with up to six barriers.

\section{Barrier Scattering}

\begin{figure*}[htb]
\centering
\includegraphics[scale=.54]{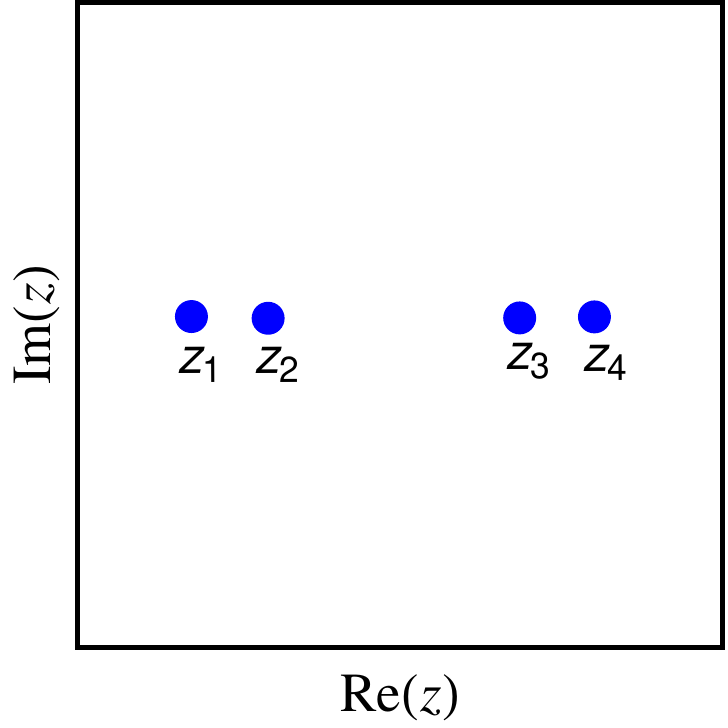}
\includegraphics[scale=.54]{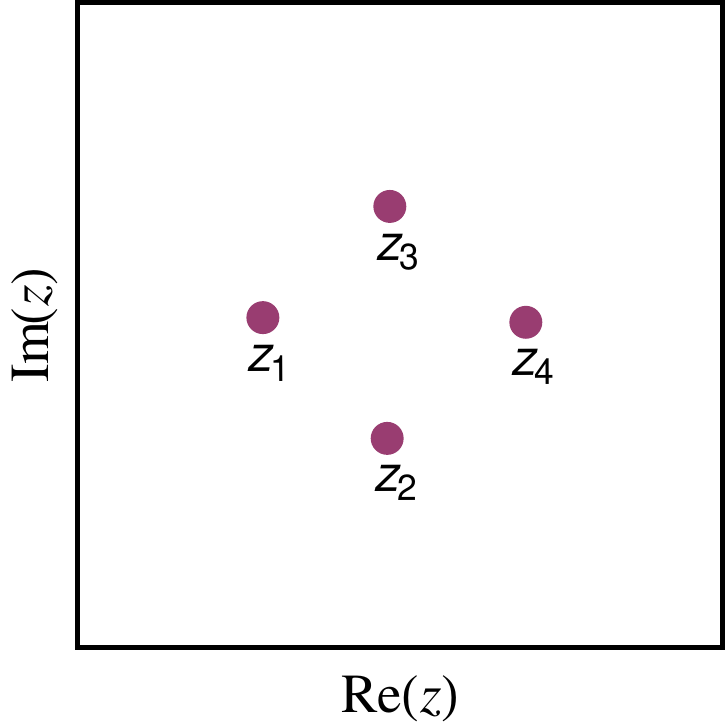}
\includegraphics[scale=.54]{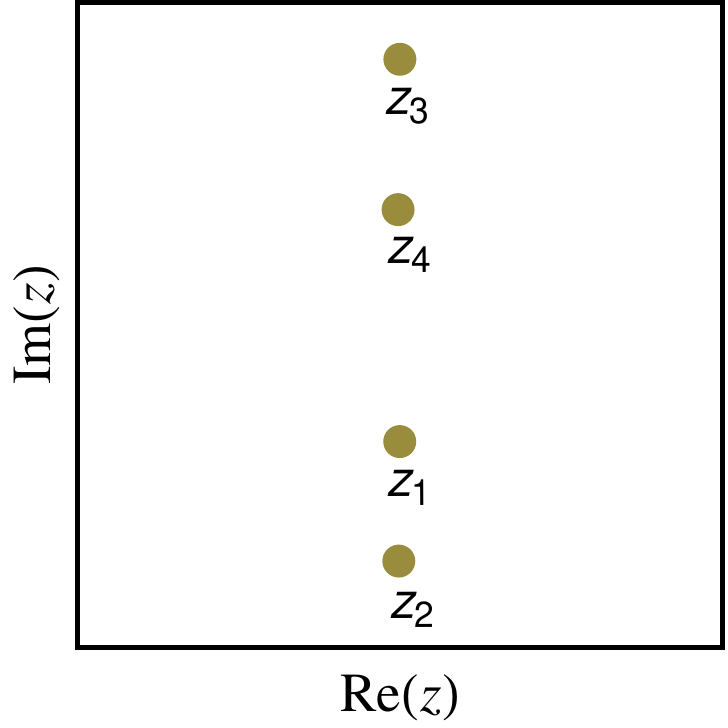}
\includegraphics[scale=.54]{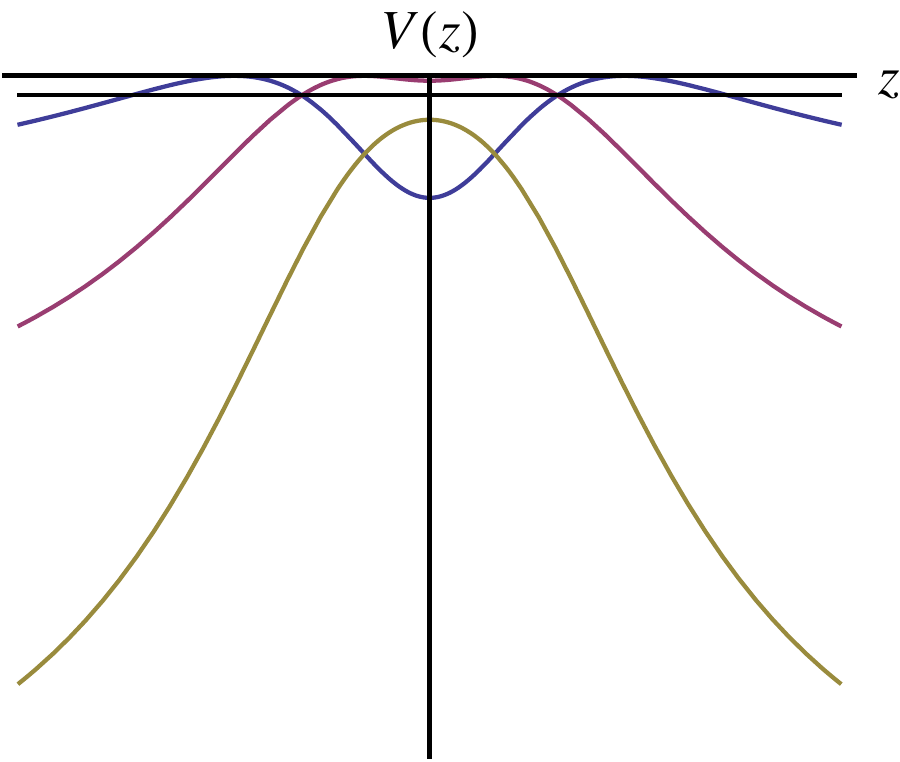}
\caption{Schematic representation of the turning points and the corresponding scattering potentials with four turning points. The horizontal line on the rightmost represents the mass level $-m^2$. Initially, in under-the-barrier scattering regime where $\omega < -m +  |q|\,\text{Max}\left[A_0(z)\right]$, all four turning points lie on the real axis. The effective potential is represented by a symmetric double barrier. The turning points of the potential well move into complex plane when  $\omega > - m +  |q|\,\text{Max}\left[A_0(z)\right]$ . Above the barrier , where   $\omega > m +  |q|\,\text{Max}\left[A_0(z)\right]$, all the points are located on the complex plane. }
\label{f1}
\end{figure*}

In the S-matrix formalism, vacuum decay rate for spinor QED is given in terms of the Bogoliubov coefficients  $\alpha(z)$ and $\beta(z)$,  which satisfy the constraint $\left|\alpha(z)\right|^2+\left|\beta(z)\right|^2=1$.  Obtaining these coefficients can be reduced to a scattering problem where one needs to solve the  Schr\"odinger-like equation:
\begin{eqnarray}
\partial^2_{z}{\phi} + \left ( p^2_{3}(z) \mp i q E(z)\right) \phi  =0  .
\label{schro1}
\end{eqnarray}
In the chiral representation of the gamma matrices, $\mp$ signs above correspond to the positive and negative helicity solutions of the Dirac equation respectively.  For our purposes it is sufficient to consider the solutions with positive eigenvalue.  In the Coulomb gauge, we consider a static source, where the electric field is given by  $\vec{E}(z)=-\partial_z A_0(z)\, \hat{\text{z}}$. The position dependent momentum is defined as $p^2_3 (z) =(\omega-q A_0(z))^2-m^2_{\perp}$ where $m^2_{\perp}=m^2+p^2_{\perp}$.  (Henceforth we set $p_{\perp}=0$). The asymptotic form of the solutions to (\ref{schro1}) is represented by  plane waves:
\begin{eqnarray}
I \,e^{-i p z} + R \, e^{i p z} \leftarrow \, \phi \, \rightarrow e^{i q z}.
\label{sol1}
\end{eqnarray}
The arrows indicate the form of the eigensolutions in the asymptotic limit $z\rightarrow\pm\infty$ respectively. The asymptotic momenta are denoted by $p$ and $q$. Note that the transmission and reflection probabilities are given by $\left|\mathcal{T}(\omega) \right|^2= 1/\left|I\right|^2$ and $\left|\mathcal{R}(\omega) \right|^2= \left|R\right|^2/\left|I\right|^2$ respectively when the transmitted amplitude is set to unity.  These probabilities are related to the vacuum decay rate which is given via $\mathcal{P}\approx \int \left|\beta(-\infty)\right|^2\, d\omega$. For this, one needs to make the identification $\left|\mathcal{T}(\omega) \right|^2/\left|\mathcal{R}(\omega) \right|^2=\left|\beta(-\infty)\right|^2 $ for the steplike barriers , such as the  Sauter field  $A_0(z)=E_0/k \tanh{k z}$, where the unitarity relation is $\left|\mathcal{R}(\omega) \right|^2-\left|\mathcal{T}(\omega) \right|^2=1$. Here, $E_0$ is the field intensity and $k$ represents the inverse width. For a symmetric double barrier potential the unitarity relation is  $\left|\mathcal{R}(\omega) \right|^2+\left|\mathcal{T}(\omega) \right|^2=1$ and we have $\left|\mathcal{T}(\omega) \right|^2=\left|\beta(-\infty)\right|^2$. The analytic form of $\left| \mathcal{T}( \omega) \right|^2$  and $\left| \mathcal{R}( \omega) \right|^2$ is obtainable for the potentials where (\ref{schro1}) is exactly solvable. For more complicated field configurations the numerical techniques becomes essential. A conventional method involves integrating the Bogoliubov coefficients on the real axis through the use of a quantum Kinetic  or a Riccati-type differential equation \cite{popov1,motto},
\begin{eqnarray}
\mathcal{P}\sim \int d\omega\int \frac{m_{\perp}E(z) }{2p^2_3}\, \text{exp}\left[2i\int^{z} p_3(z')\,  dz'\right]  dz .
\label{ric}
\end{eqnarray}
In spatially inhomogeneous fields  computation of (\ref{ric}) becomes exceedingly difficult because in principle, there can be an arbitrary number of turning points i.e zeros of $p(z)$ on the real axis. More specifically, for supercritical potentials ($|q|\text{Max}\left[A_0(z)\right] > 2 m$) there can be at least a single pair of zeros on the real axis in under-the-barrier scattering regime where $\omega < m + |q|\,\text{Max}\left[A_0(z)\right]$. To see this, it is useful to define the effective scattering potential as $V(z)=-(\omega-q A_0(z))^2=-p^2_0(z)$. The turning points are given by $V(z_{\text{tp}})=-m^2$. These points  move into complex $z$ plane to form complex conjugate pairs as the parameter $\omega$ is increased (Fig. \ref{f1}). The value $\omega =m+ |q|\,\text{Max}\left[A_0(z)\right]$ represents the barrier top. For energies above the barrier, all the turning points are located on the complex plane and (\ref{ric}) can be integrated without difficulty.

\subsection{Semiclassical formalism and under-the-barrier scattering}

In under-the-barrier scattering regime, the effective potential with an arbitrary shape may contain multiple barriers. Transmission and reflection amplitudes can be obtained in the framework of semiclassical approximation. For this, one needs to analytically continue the WKB solutions across the barriers in the scattering potential. The WKB ansatz for (\ref{schro1}) with positive helicity is given as
\begin{eqnarray}
&&\phi = a \,\lambda^+\, e^{i \int^z p_3(z')} + b\,\lambda^-\, e^{-i \int^z p_3(z')} ,\nonumber\\
\small
&&\lambda^{\pm}= \left[p_3(z)\left(p_0(z)\pm p_3(z)\right)\right]^{-1/2}.
\label{ans}
\end{eqnarray}
Here, $a$ and $b$ are the constant coefficients of incoming and outgoing solutions in the region of interest.
The coefficients which are located on the left and on the right side of a single potential barrier must be related in a way that the value of the Dirac current along $z$ remains unchanged. We may write the conserved current along $z$ in terms of the positive and negative helicity solutions such that $J^3=\int d\omega \sum_{s} j^3_s$ where $s$ denotes the helicity index. The positive helicity current $j^3_+$  can be written as
\begin{eqnarray}
j^+_3&=&p^2_3(z)\left| \phi \right|^2 + \left| \partial_z\phi \right|^2 + i p_0(z) \left(\phi\partial_z\phi^*-\phi^*\partial_z\phi\right).
\label{cur}
\end{eqnarray}
Upon substitution of (\ref{ans}) into (\ref{cur}) we obtain
\begin{eqnarray}
j^+_3 &=& \left(1+ \frac{E^2(z)}{2p^4_3(z)}\right)(\left|a\right|^2-\left|b\right|^2)+\frac{E^2(z)p_0(z)}{2p^5_3(z)}(\left|a\right|^2+\left|b\right|^2)\nonumber\\
&+&\frac{E^2(z)m}{4p^5_3(z)}\left(a\, b^*\, e^{2i\int^z p_3(z')}+a^*\, b \,e^{-2i\int^z p_3(z')}\right).
\end{eqnarray}
The energy is of the order $p_0(z) \sim p_3(z) $, thus in the parameter regime where $\partial_3p_3(z)/p^2_3(z) \ll 1$ we may write
\begin{eqnarray}
 j^+_3\approx \left|a\right|^2-\left|b\right|^2.
\label{wuni}
\end{eqnarray}
We label the constant multipliers of (\ref{ans}) by $a^l$ and $b^l$ to the left of the barrier and by $a^r$ and $b^r$ to the right. The set coefficients $a^l$ and $b^l$ are related to the $a^r$ and $b^r$ by using a set of analytic continuation rules across potential barrier\cite{heading}. Doing so we may write the solutions as
\begin{eqnarray}
\phi^l = a^l \,\lambda^+\, e^{i \int^z_{z_1} p_3(z')} + b^l \,\lambda^-\, e^{-i \int^z_{z_1} p_3(z')} ,\nonumber\\
\phi^r = a^r \,\lambda^+\, e^{i \int^z_{z_2} p_3(z')} + b^r \,\lambda^-\, e^{-i \int^z_{z_2} p_3(z')},
\end{eqnarray}
where $z_1$ and $z_2$ are the turning points of the barrier ($z_2>z_1$). The coefficients $a_r$ and $b_r$ are
\begin{eqnarray}
a^r &=& e^{K}\left(S_1 a^l + b^l\right),\nonumber\\
b^r & =& -e^{-K}\left( (1+S_1S_2)a^l + S_2 b^l \right).
\label{lr}
\end{eqnarray}
Here $S_1$ and $S_2$ are the Stokes constants. The exponent is  given as $K=-i\int^{z_2}_{z_1}p_3(z) dz  \,\,(K>0)$. The Stokes constants can be determined up to a phase by imposing
\begin{eqnarray}
\left|a^l\right|^2-\left|b^l\right|^2=\left|a^r\right|^2-\left|b^r\right|^2
\end{eqnarray}
which yields
\begin{eqnarray}
S_1=i\sqrt{e^{2K} +1} e^{-K}e^{i\varphi},\,\, S_2=i\sqrt{e^{2K} +1} e^{K}e^{-i\varphi}
\end{eqnarray}
It is convenient to write (\ref{lr}) in the matrix form,
\begin{eqnarray}
\begin{pmatrix}
a^r \\
b^r
 \end{pmatrix}&=&
\begin{pmatrix}
 e^{K}S_1 & e^{K} \\
  -e^{-K}(1+S_1 S_2) &   -e^{-K}S_2
 \end{pmatrix}\begin{pmatrix}
  a^l \\
  b^l
 \end{pmatrix}.
\label{conn1}
\end{eqnarray}
For the set of eigensolutions in (\ref{sol1}), we may invert the matrix in (\ref{conn1}) and write
\begin{eqnarray}
\begin{pmatrix}
a^l \\
b^l
 \end{pmatrix}&=&
\begin{pmatrix}
  -i \sqrt{e^{2K}+1}e^{-i\varphi} & -e^{K} \\
  -e^{K} &   i \sqrt{e^{2K}+1}e^{i\varphi}
 \end{pmatrix}\begin{pmatrix}
  a^r \\
  b^r
 \end{pmatrix}
\label{conn2}
\end{eqnarray}
The connection matrix $M$ in the above equation satisfies $\text{det}\,M=1$. The phase $\varphi$ is associated with the Stokes' constant. This can be determined by matching the exact solutions of (\ref{schro1}) for the parabolic barrier with the WKB ansatz in (\ref{ans}). In the first-order approximation, we have \cite{froman1}
\begin{eqnarray}
\varphi&=&\text{arg}\left[\Gamma\left(\frac{1}{2}+i \frac{K}{\pi}\right)\right]-\frac{K}{\pi}\left(\log{\left|\frac{K}{\pi}\right|}-1\right).
\end{eqnarray}
The coefficient of the transmitted wave is represented by the exponentially large part of the solution $\phi^r$. After fixing this coefficient, one can get the transmission and reflection probabilities for a multi-barrier potential by the successive application of (\ref{conn2}). Regardless of the unitarity relation, we may write the vacuum decay rate in an arbitrary external field  in the form
$\left|\beta(-\infty)\right|^2= 1/(1+ f(\omega))$  where $f(\omega)$ consists of phase integrals.   The application of (\ref{conn2})  for an external field with a single pair of turning points in the scattering potential (Sauter field) yields $f^1(\omega) = e^{2K}$. The accuracy of the semiclassical formula is well assured if $E_0/k \gg m$.  This also corresponds to the parameter regime where the amplitude becomes supercritical.

\begin{figure}[htb]
\includegraphics[scale=0.61]{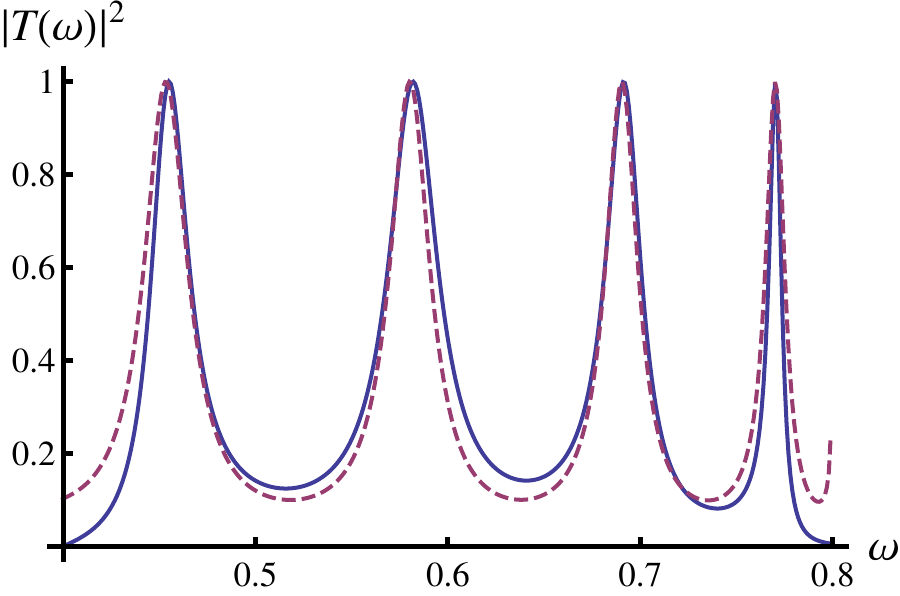}
\caption{The transmission probability for the Wood-Saxon potential in (\ref{wood}). The exact result is given in \cite{wood} and is shown by solid blue curve. The dashed red curve shows the result of approximate formula in (\ref{double}). The parameters are chosen in accordance with \cite{wood}: $m=0.4,\, d=10,\, k=25$ and $W=1.2$ }
\label{f2}
\end{figure}

Obtaining a semiclassical formula for the vacuum decay rate becomes increasingly tedious as the number of barriers increases, yet the calculation is straightforward. Here,  we carry out the calculation for a double barrier potential with four turning points$(z_1..z_4)$. In the Appendix we simply give out the rules  for an arbitrary potential with $n$ barriers. Before proceeding we label $\phi$  such that $\phi_j^l$ and $\phi^r_j$ represent the solutions to the left and to the right of the $j$\,th barrier respectively.  Fixing the coefficient of the transmitted wave $\phi^r_2$ to the left of the second barrier as $a^r_2$, we may write $\phi^r_2= a^r_2 \lambda^+ e^{i\int^z_{z_4} p(z')}$. Applying the connection matrix we get
\begin{eqnarray}
\phi^l_2 = a^r_2 \,M_2^{11} \lambda^+ e^{i\int^z_{z_3} p(z')} + b_2^r\, M_2^{21} \lambda^- e^{-i\int^z_{z_3} p(z')}.
\end{eqnarray}
The solution to the right of the first barrier is obtained by simply continuing the above solution across the potential well,
\begin{eqnarray}
\phi_2^l \equiv \phi_1^r &=&a_2^r \,M_2^{11} e^{-i L} \lambda^+ e^{i\int^z_{z_2} p(z')} \nonumber\\
&+& b_2^r\, M_2^{21} e^{i L} \lambda^- e^{-i\int^z_{z_2} p(z')},
\end{eqnarray}
where $L=i\int^{z_3}_{z_2} p(z') $. Upon acting $M_1$ on $\phi_1^r$ one gets
\begin{eqnarray}
\phi^l_1 &=& a^l_1 \lambda^+ e^{i\int^z_{z_1} p(z')} +  b^l_1 \lambda^- e^{-i\int^z_{z_1} p(z')},\nonumber\\
a^1_l  &=& (M_1^{11} M_2^{11} e^{-i L} + M_1^{12} M_2^{21} e^{i L})\, a^r_2, \nonumber\\
b^1_l  &=& (M_1^{21} M_2^{11} e^{-i L} + M_1^{22} M_2^{21} e^{i L})\, a^r_2.
\end{eqnarray}
Setting $a^r_2=1$, the transmission and reflection amplitudes are simply $\mathcal{T}=1/a^1_l$ and $\mathcal{R}=a^1_l/b^1_l$. By virtue of the unitarity relation  $\left|\mathcal{R} \right|^2+\left|\mathcal{T}\right|^2=1$, we may write $\left|\beta(-\infty) \right|^2=\left|\mathcal{T} \right|^2$ where
\begin{eqnarray}
\left|\mathcal{T}(\omega) \right|^2&=& \frac{1}{(1+ f^2(\omega))}\nonumber\\
f^2(\omega)&=& e^{2K_1}+e^{2K_2}+2\,e^{2K_1+2K_2}-2\, e^{K_1+K_2}\nonumber\\
&&\times\sqrt{e^{2K_1}+1}\sqrt{e^{2K_2}+1}\cos{\left(2L+\varphi_1+\varphi_2\right)}.\nonumber\\
\label{double}
\end{eqnarray}
In the beginning, had we chosen the coefficient of $\phi^r_2$ as $b^r_2$, we would have gotten $\left|\mathcal{R} \right|^2-\left|\mathcal{T}\right|^2=1$. This would not affect the final result since the ratio $\left|\mathcal{T} \right|^2/\left|\mathcal{R} \right|^2$ would  precisely give  $1/(1+ f^2(\omega))$. In Fig 2. we use the exact result for the transmission probability in the symmetric double barrier Woods-Saxon potential\cite{wood}
\begin{equation}
A_0(z) =W\left(\frac{\theta(-z)}{1+e^{-k(z+d)}}+\frac{\theta(z)}{1+e^{k(z-d)}}\right),
\label{wood}
\end{equation}
and compare it  with (\ref{double}). Here, $d$ is the offset and $W$ is the amplitude.  Despite the fact that (\ref{wood}) has a cusp at $z=0$, it can be smoothed out for the values $k \gg d$. Fig. \ref{f2} shows the transmission resonances for the Woods-Saxon potential in such parameter regime.  The agreement between the exact result and semiclassical formula is  remarkable; the WKB result accurately yields the widths and the positions of the resonance peaks.

For $n$ pairs of turning points we may get a useful formula for $f^n(\omega)$ (see Appendix for the details) which, in the leading order, can be written as :
\begin{eqnarray}
f^n(\omega)&\approx& 4^{n-1} \prod^{n}_{i=1}  e^{2K_{i}}\prod^{n-1}_{j=1}\sin^2 {\tilde{L}_{j}},\nonumber\\
\tilde{L}_{j}&=&L_j+\varphi_j/2+\varphi_{j+1}/2,
\label{npair}
\end{eqnarray}
where the index $i$ refers to turning point pair and the phase $L_j$ connects the two consecutive pairs. It is understood that $\varphi_j$ represents the phase for each barrier. Note that the lower-order exponentials can be safely neglected if the resonance peaks are not too close to each other. We will make use of (\ref{npair}) in the next section to discuss the effect of multiple wells on the transmission resonance.

\subsection{Multiple well potentials and transmission resonances}
 \begin{figure*}
\includegraphics[scale=0.62]{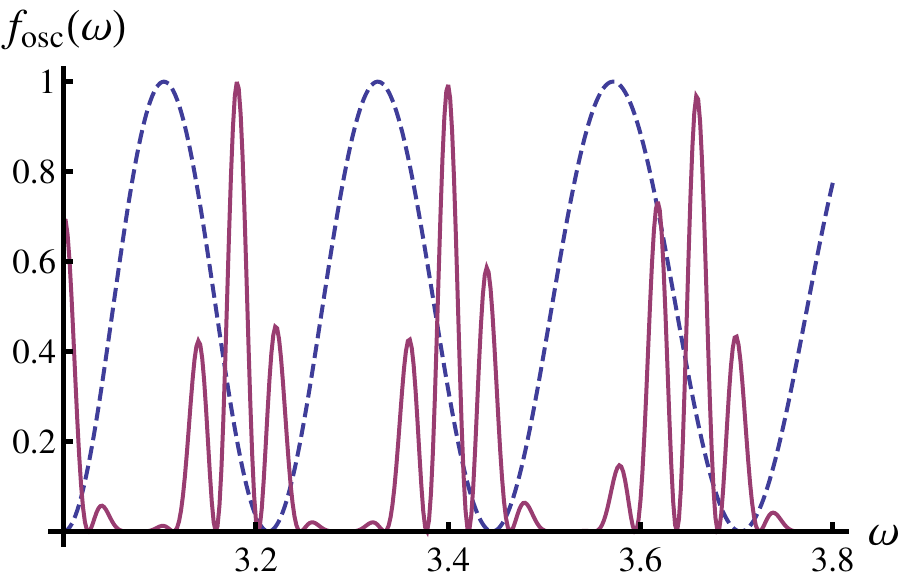}
\includegraphics[scale=0.62]{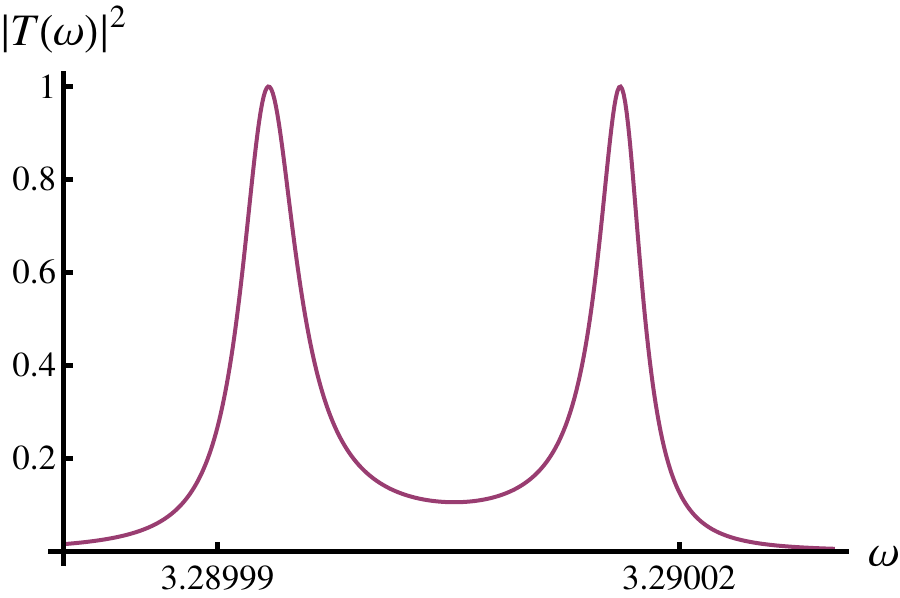}
\includegraphics[scale=0.62]{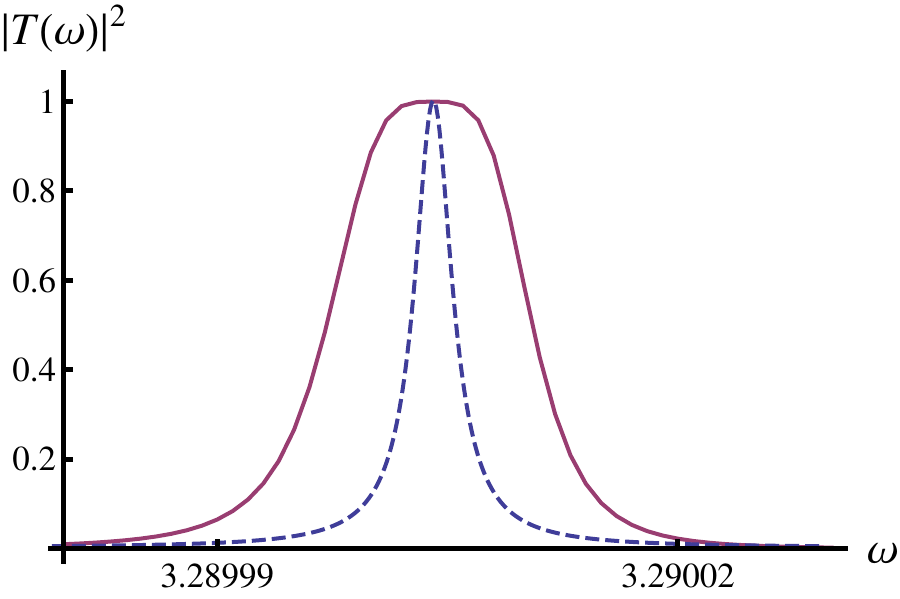}
\caption{The oscillatory part of $f^4(\omega)$ in (\ref{npair}) for the single well (dashed,blue) and the triple well (solid,red)  configurations in under-the-barrier scattering regime. Remaining figures display resonance peaks of the triple-well field (solid,red). The field parameters are ($m=1$): $E_0=0.5$, $k=0.1 $ and $d=41.07076$ (left, middle) and $d=41.07077$ (right). The dashed blue curve (right) represents the largest resonance peak of the single well configuration (located at $\omega=3.7087$) for comparison.}
\label{f3}
\end{figure*}

\begin{figure}[htb]
\includegraphics[scale=0.61]{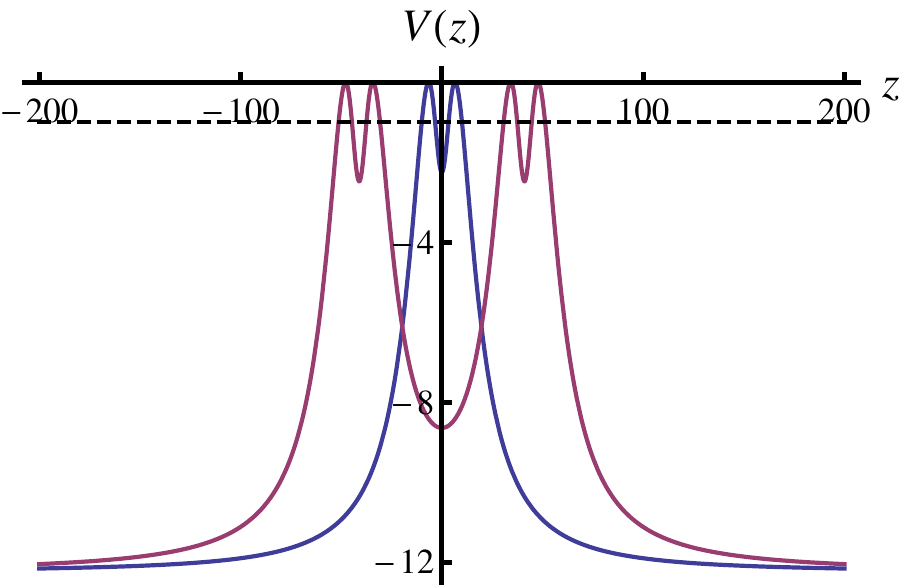}
\caption{The effective scattering potentials for the chosen external fields in the text. The dashed line shows the normalized mass level $-m^2=-1$ where we have set $p_{\perp}=0$. The field parameters are given in terms of the normalized mass as: $E_0=0.5, \,k=0.1$, \,$d =41.07$ and $\omega=3.5 $. }
\label{f4}
\end{figure}
In order to work with a time-independent electric field in vacuum we consider a static source. In the simple case where the external field is constant, the source could be given as seperated and oppositely charged parallel plates. A simple setup to achieve more complicated field configurations can be imagined as an array of charged rings in static equilibrium, which is centered along the $z$ axis. In the following, we assume the \textit{shape} of the electric field  in the vacuum region  can be modeled by an effective potential $A_0(z)$, which is represented by the  parameters  $E_0$ and $k$. To see the effect of multiple wells in the effective scattering potential $V(z)$, we write $A_0(z)$ as:
\begin{eqnarray}
A_0(z)=-\frac{E_0}{k}\sum_{i} \frac{1}{1+k^2 (z-d_i)^2}.
\label{nbarrier}
\end{eqnarray}
In the following we compare the transmission resonances of the single well configuration,
\begin{eqnarray}
A_{0}(z)=-\frac{E_0/k}{1+k^2 z ^2},
\label{1well}
\end{eqnarray}
with the  triple-well configuration(see Fig. \ref{f4}) which is given as
\begin{eqnarray}
A_{0}(z)=-\frac{E_0/k}{1+k^2 (z-d) ^2}+\frac{E_0/k}{1+k^2 (z+d) ^2}
\label{3well}
\end{eqnarray}
In Fig. \ref{f3}, we plot the phase terms in $f(\omega)$ for the chosen electric fields by using (\ref{npair}). In general,  the increasing number of oscillatory terms opens up new channels for the transmission resonance. Some of the resonance peaks are strongly suppressed due to increasing number of the exponential terms. On the other hand, when zeroes of the oscillatory terms overlap at specific channels, resonance widths get amplified. To see where such overlapping may occur, semiclassical formula (\ref{npair}) is particularly instructive. Upon inclusion of the lower-order exponentials in the definition, specific resonance widths can be enhanced. For the triple-well potential considered here, this could be achieved by adjusting the value of $d$,  which determines the width of the larger well in the middle. This adjustment brings two neighboring resonance peaks together as depicted in Fig. \ref{f3}.   Here, It should be emphasized that the number of such closely spaced resonance peaks increases with the number of the wells and therefore the broadening effect is directly related to  the shape of the effective scattering potential. Moreover, the resonance width shows extreme sensitivity to the  field parameters. The adiabaticity parameter which is defined as $\gamma=\frac{m k}{q E_0}$, is especially useful in seeing this. If one keeps the ratio $k/E_0$ fixed by changing $E_0$ and $k$ the same amount, positions of the resonance peaks for a symmetric double barrier potential remain almost intact, but the change in the resonance widths is dramatic.

\subsection{Above-the-barrier-scattering and Riccati qquation}

As the value of $\omega$ is increased the turning points start moving into complex plane. Generally, this happens in a fashion that the turning points which determine the phases $L_j$ get close, coalesce and then they get shifted to complex plane in the form of complex conjugate pairs (Fig.1). During this process, $L_j$ and therefore phase terms in $f(\omega)$ show no discontinuity. We may still use the connection matrix in  (\ref{conn2}) as long as $e^{2K}\gg 1$ and $L_j \gg \varphi_j/2+\varphi_{j+1}/2$. As one keeps increasing $\omega$, turning points that govern the exponential terms will also approach to each other and, eventually the WKB approximation will break down. Close to the barrier top, the connection formula may be remedied by going to higher orders in the approximation\cite{froman2,bender}. This requires closed form of the phase integrals which is obtained by using a suitable contour. The next leading-order phase integrals yield complicated expressions in terms of the elliptic functions for the potential in (\ref{nbarrier}), but as the energy gets closer to the barrier top, one needs in general the higher order terms to get the desired accuracy. This makes the whole analysis cumbersome, and in the case of more complicated field configurations, closed form expressions for the integrals may not even exist.

\begin{figure}[htb]
\includegraphics[scale=0.85]{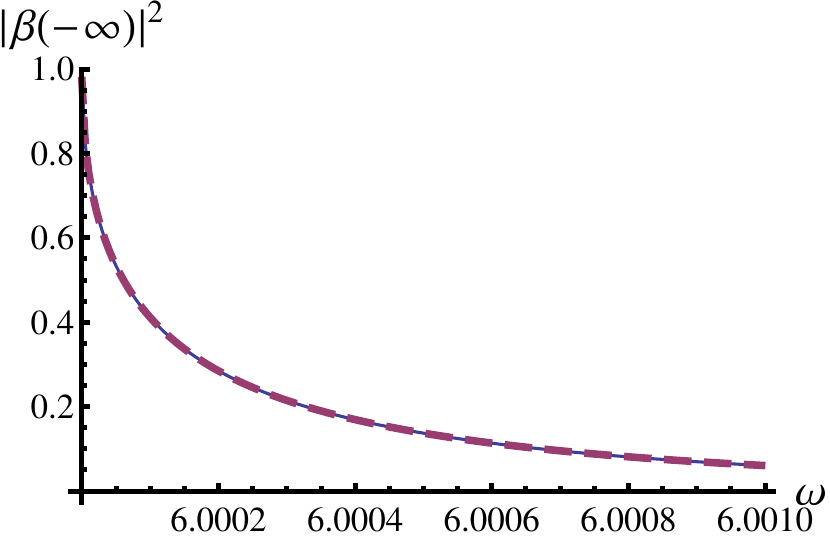}
\caption{Vacuum decay rate for the Sauter field. The analytic result was given in \cite{niki1}. The dashed red curve is obtained by numerical integration of (\ref{couple}). We perform the computation in over-barrier-scattering regime ($\omega > m + E_0/k $) which is located beyond the elastic scattering regime ( $E_0/k -m \leq \omega \leq  E_0/k + m $). The field parameters are $m=1,\, E_0=.5, \, k=.1$. The integration can satisfactorily be optimized to yield accuracy  performance up to 8-9 significant figures.}
\label{f5}
\end{figure}

Above the barrier, all the turning points form complex conjugate pairs. This is reminiscent  of scattering in a time-dependent electric field. In this regime the Bogoliubov coefficients can be computed  numerically via WKB-like ansatz:
\begin{eqnarray}
\phi = a(z) \,\lambda^+\, e^{i \int^{z} p_3(z')} + b(z)\,\lambda^-\, e^{-i \int^{z} p_3(z')},
\label{ans2}
\end{eqnarray}
where now the coefficients $a$ and $b$ depend on the position. Inserting (\ref{ans2}) into (\ref{schro1}) and imposing the consistency relation
\begin{eqnarray}
\partial_z\phi =ip_3(z) ( a(z) \,\lambda^+\, e^{i \int^{z} p_3(z')} - b(z)\,\lambda^-\, e^{-i \int^{z} p_3(z')}),\nonumber
\end{eqnarray}
we have
\begin{eqnarray}
\partial_z a (z) &=&\frac{b(z) E(z) m_{\perp}}{2p_3^2(z)}\,e^{-2 i \int^{z} p_3(z') d z'},\nonumber\\
\partial_z b (z) &=&\frac{a (z) E(z) m_{\perp}}{2p_3^2(z)}\,e^{2 i \int^{z} p_3(z') d z'}.
\label{couple}
\end{eqnarray}
The numerical integration can be performed after setting the coefficient of the transmitted wave to unity. The asymptotic values of the coefficients are related to Bogoliubov coefficients as $|\alpha (-\infty)|^2=1/\left|a(-\infty)\right|^2$ and  $|\beta (-\infty)|^2=\left|b(-\infty)\right|^2 /\left|a(-\infty)\right|^2$. In Fig. \ref{f5} we compare the analytical result for the Sauter field with the numerical integration.  Our numerical investigation with several electric field profiles reveals that at the barrier top, $ |\beta (-\infty)|^2$ starts with a value of the order of unity and  decays sharply.  The use of (\ref{couple}) for various values of $E_0$ and $k$ shows the width of the decay is very sensitive to $k$, where $E_0$ in general shifts the location of the barrier top.  No resonance peaks are expected in this region nevertheless,  relative locations of the turning points may allow for observable interference effects \cite{dumlu2}. 

\section{Conclusion}
In this paper, we have investigated the energy spectrum of the positrons produced by a spatially inhomogeneous external field with subcycle structure. We have given approximate formulas for the transmission probability in under-the-barrier-scattering regime, where we have found that the number of resonance peaks increases with the increasing number of wells in the effective scattering potential, and certain resonance peaks may be broadened by fine-tuning of the width of the wells. This could be relevant for lowering the time interval needed to keep the external field on, before a positron is spontaneously emitted. In above-the-barrier scattering regime numerical integration of the Bogoliubov coefficients can be performed without difficulty. Close to the barrier top,  transmission probability is at the order of unity and it sharply falls off, where the decay width is determined by the inverse width of the external field. In this respect, a comparison of the total vacuum decay probability for a purely space-dependent and a time dependent field with the same field parameters might be appealing.  As far as both space and time dependent electric fields are concerned,  the interplay between the temporal and the spatial profile of the external field may become particularly important, when the resonant tunneling time becomes comparable to the timescale set by the laser frequency. 
\bigskip

This work was supported by T\"{U}B\.{I}TAK through Grant No. 112C008. I would like to thank T. Birol, G. Dunne, and B. Tekin for valuable comments and suggestions.

\section{Appendix: Transmission Probability for an External Potential with an Arbitrary Number of Barriers}

We formulate the general form of $f^n(\omega)$ inductively through the procedure outlined in Sec. II and, give the rules to obtain all the terms in $f^n(\omega)$ for $n$ pairs of turning points. The total number of terms of in $f^n(\omega)$ is given by $4^{n-1}$ so we will not be writing the explicit form $f(\omega)$ for a  higher number of turning point pairs; nevertheless  it is useful to go one step further and write the result for three pairs of turning points.
\begin{widetext}
\begin{eqnarray}
f^3(\omega) &=& (e^{2K_1}+e^{2K_2}+e^{2K_3})+ 2 (e^{2K_2+2K_1}+e^{2K_3+2K_1}+e^{2K_3+2K_1}) + 4\, e^{2K_3+2K_2+2K_1}\nonumber\\
&&-2 \,(1+2e^{2K_3} )\sqrt{e^{2K_2}+1}\, e^{K_2}\sqrt{e^{2K_1}+1}\,e^{K_1}\cos{2\tilde{L}_1}-2 \,(1 + 2e^{2K_1})\sqrt{e^{2K_3}+1}\, e^{K_3}\sqrt{e^{2K_2}+1}\,e^{K_2}\cos{2\tilde{L}_2}\nonumber\\
&&+2\, \sqrt{(e^{2K_1}+1)(e^{2K_3}+1)}\,e^{K_1+K_3}\left((1+e^{2K_2})\cos{(2\tilde{L}_2+2\tilde{L}_1)}+ \,e^{2K_2}\cos{(2\tilde{L}_2-2\tilde{L}_1)}\right).
\end{eqnarray}
\end{widetext}
The general form of $f^n(\omega)$ is composed of two main parts $f^n_e(\omega)$ and $f^n_o(\omega)$, the first of which is given by purely exponential terms whereas the second contains oscillatory terms. Our first observation is\\

(I) : The general form  of $f_e(\omega)$  can be written as the sum:
\begin{widetext}
\begin{eqnarray}
f^n_e(\omega) &=&  \sum_{\mathclap{i}} e^{2K_{i}} + 2\sum_{\mathclap{i_1>i_2}} e^{2K_{i_1}+2K_{i_2}}
+ 2^2 \sum_{\mathclap{i_1>i_2>i_3}} e^{2K_{i_1}+2K_{i_2}+2K_{i_3}}\,\,\,......
+2^{n-2} \sum_{\mathclap{i_1>..>i_{n-1}}}e^{2K_{i_1}..+2K_{i_{n-1}}} + 2^{n-1} e^{\sum^{n}_{i}2K_{i}}.
\label{ge}
\end{eqnarray}
\end{widetext}
where summation over each index $i_{j}$ is implied. The exponents that appear in (\ref{ge}) represent all the possible distinct $j$-tuple sums, which are constructed from the the set $s^n_{K}=\{K_1,K_2...K_n\}$, where $j$ runs from 1 to $n$.

In the remaining part $f^n_o(\omega)$, each term has a oscillatory part $\cos{(Y(\tilde{L}))}$,  which is multiplied by an exponential factor. Here, the argument  $Y(\tilde{L})$ can be  composed of any the distinct $m$-tuple combination of $\tilde{L}$'s which is constructed from the set $s^{n}_{\tilde{L}}=(2 \tilde{L}_{n-1},  ..,  2 \tilde{L}_{2}, 2 \tilde{L}_{1})$, where $1 \leq m \leq n-1$  (we assume there are $n-1$ wells in an effective potential with $n$ barriers.)  For instance, the terms in $f_o(\omega)$ for $n=4$ and $m=2$ can be written as
\begin{widetext}
\begin{eqnarray}
f^4_o(\omega) &=&...+ X^{2,1^+}\cos{(2\tilde{L}_2+2\tilde{L}_1}) + X^{2,1^-}\cos{(2\tilde{L}_2-2\tilde{L}_1})
+ X^{3,2^+}\cos{(2\tilde{L}_3+2\tilde{L}_2})\nonumber\\
&& +X^{3,2^-}\cos{(2\tilde{L}_3-2\tilde{L}_2})
+ X^{3,1^+}\cos{(2\tilde{L}_3+2\tilde{L}_1})
+ X^{3,1^-}\cos{(2\tilde{L}_3-2\tilde{L}_1})
...
\end{eqnarray}
\end{widetext}
The coefficients $X^{2,1^+}, X^{2,1^-},..., X^{3,1^-}$ represent the exponential factors that are fixed by by the argument of the cosine they multiply. Note that the argument of the cosine includes all the possible sign permutation of $\tilde{L}$'s except for the first entry.  We may  write the general form  $f_o(\omega)$ as:
\begin{widetext}
\begin{eqnarray}
f^n_o(\omega)&=&\sum^{n}_{\mathclap{i=1}} X^{i}\cos{ (2\tilde{L}_i)}
+\sum_{\mathclap{i_1>i^{s}_2}}X^{i_1,i^{s}_2}\cos{ (2\tilde{L}_{i_1} + 2\tilde{L}_{i^{s}_2})}+
\sum_{\mathclap{i_1>i^{s}_2>i^{s}_3}} X^{i_1,i^{s}_2,i^{s}_3}\cos{ (2\tilde{L}_{i_1} + 2\tilde{L}_{i^{s}_2} + 2\tilde{L}_{i^{s}_3})}....\nonumber\\
&&....+\sum_{\mathclap{i_1>..>i^{s}_{n-2}}}X^{i_1,..,i^{s}_{n-2}}\cos{ (2\tilde{L}_{i_1}....+ 2\tilde{L}_{i^{s}_{n-2}})} + +\sum_{\mathclap{i_1>..>i^{s}_{n-1}}}X^{i_1,..,i^{s}_{n-1}}\cos{ (2\tilde{L}_{i_1}....+ 2\tilde{L}_{i^{s}_{n-1}})}
\label{fe}
\end{eqnarray}
\end{widetext}
Here, we have used the double index $i^{s}_{j}$ where $s$ fixes the sign in front of each $L_{i_j}$ that appears in the argument. Each sum is to be performed over $(i_{1}, i_j)$ and $s$ as well to include all the sign permutations. In the following we show how to determine the factors in the summations. For this, we consider a generic argument $Y^{m}(\tilde{L})$, which is constructed by using an arbitrary subset $s^{m}_{\tilde{L}}$ of  $s^n_{\tilde{L}}$ where $m$ is the total number of elements in the set. We denote the corresponding factor multiplying $\cos{(Y^{m}(\tilde{L}))}$ as $X^{m}_n(K)$. The form $X^{m}_n(K)$ can be determined by the following steps:\\

(II) : The sign of $X^{m}_n(K)$ is given by $(-1)^{m}$.\\

(III): For any isolated $\tilde{L}_{k}$ in $Y^m({\tilde{L}})$ such that there are no neighboring terms $\tilde{L}_{k-1}$ and $\tilde{L}_{k+1}$,  $X^{m}_n(K)$ gains a factor:
\begin{eqnarray}
\tilde{L}_{k}\rightarrow e^{K_{k}+K_{k+1}} \sqrt{e^{2K_{k}}+1}\sqrt{e^{2K_{k+1}}+1}.\nonumber
\end{eqnarray}

(IV) : If there are neighboring elements such as $L_{k}$ and $\tilde{L}_{k+1}$, the factors they introduce have a common exponent $K_{k+1}$.  The form of the composite factor where the common exponent appears depends on the relative signs of the  $\tilde{L}_{k}$ and $\tilde{L}_{k+1}$. There are two possible cases. In the first case $\tilde{L}_{k}$ and $\tilde{L}_{k+1}$ carry the same sign in the argument and the common factor is given by $e^{2K_{k+1}}+1$. If the signs are opposite, one has $e^{2K_{k+1}}$. We may write this as
\begin{eqnarray}
\pm \tilde{L}_{k}\pm  \tilde{L}_{k+1}& \rightarrow & (e^{2K_{k+1}}+1)e^{K_k+K_{k+2}}\nonumber\\
&\times&\sqrt{e^{2K_{k}}+1}\sqrt{e^{2K_{k+2}}+1},\nonumber\\
\pm \tilde{L}_{k}\mp \tilde{L}_{k+1} &\rightarrow & \,e^{2K_{k+1}}e^{K_k+K_{k+2}} \sqrt{e^{2K_{k}}+1}\sqrt{e^{2K_{k+2}}+1}.\nonumber
\end{eqnarray}
This applies to any array $\tilde{L}_a= \tilde{L}_{k}+\tilde{L}_{k+1}-\tilde{L}_{k+2}+...$ with an arbitrary number of neighboring terms. According to III and IV, we may infer the total number of exponents and the square root terms that are introduced to $X^{m}_n(K)$ by the argument $Y^{m}(\tilde{L})$. We will use this information to completely characterize $X^{m}_n(K)$ by the argument.

Suppose in the argument $Y^{m}(\tilde{L})$ there are $n'$ isolated $\tilde{L}_{k}$ s and  $m'$ arrays of arbitrary length. Total number of terms due to arrays can be written as
\begin{eqnarray}
\sum^{m'}_{i} m'_{i}= m-n'.
\end{eqnarray}
where $m'_i$ is the number of elements in $i$th array $\tilde{L}^{i}_a$. Total number of distinct exponents introduced by all $\tilde{L}^{i}_a$ is simply
\begin{eqnarray}
\sum^{m'}_{i} m'_{i}+1= m-n'+m'.
\end{eqnarray}
Taking into account $2n'$ exponents introduced by the isolated terms, total number of  exponents fixed by the argument $Y^m(\tilde{L})$ is $m_c=m+n'+m'$. According to III and IV, the total number of square root terms that appear in $X^{m}_n(K)$ is 2$n'$+2$m'$.\vspace{.2cm}

(V) :  Every pair of square root terms that appears in  $X^{m}_n (K)$ is accompanied by a multiplicity factor  2. The total multiplicity factor is $2^{n'+m'}$\vspace{.2cm}

Note that  $m_c < n$ in general so the exponents fixed by III and IV cover only a subset $s^{m_c}(K)$ of $s^{n}_K$. Let us call the portion of $X^{m}_{n}(K)$ fixed by I-V as $X^{m_c}_{n}(K)$. We define the complementary part $X^{c}_{n}(K)$ such that
\begin{eqnarray}
X^{m}_{n}(K)=X^{m_c}_{n}(K) X^{c}_{n}(K).
\end{eqnarray}\vspace{.1cm}

(VI) : The complementary part $X^{c}_n(K)$ emerges as the sum
\begin{widetext}
\begin{eqnarray}
X^{c}(K) &=& 1 + 2\sum_{\mathclap{i}} e^{2K_{i}} + 2^2\sum_{\mathclap{i_1>i_2}} e^{2K_{i_1}+2K_{i_2}} + 2^3 \sum_{\mathclap{i_1>i_2>i_3}} e^{2K_{i_1}+2K_{i_2}+2K_{i_3}}
...
+ 2^{n-m_c} e^{\sum^{n-m_c}_{i}2K_{i}}
\label{for}
\end{eqnarray}
\end{widetext}
where
\begin{eqnarray}
(K_i, K_{i_{j}}) &\in&  s^{c}_{K}, \,\,\, 2 \leq j \leq n-m_c\nonumber\\
s^{c}_{K}&=&s^{n}_{K} -  s^{m_c}_{K}
\end{eqnarray}
Similar to (\ref{fe}), the exponents in each sum above appear as all the distinct $j$-tuple sums from the set $s^{c}_{K}$. Using II-VI one can in principle completely determine $X^{m}_n(K)$. The steps I-VI presented here help us to consistently reproduce all the elements of $f(\omega)$ for any $n$ without having to use the connection matrix for an arbitrary scattering potential.

We will now show $f(\omega)$ can be greatly simplified if one considers only the leading-order terms. This requires organizing the terms which are multiplied with the leading-order exponential $e^{\sum^{n}_{i} 2K_{i}}$.  For this, we  only take the leading-order term in any root term that appear in $X^{m}_{n}(K)$ such that
\begin{eqnarray}
\sqrt{e^{2K_{k}}+1} &=& e^{K_{k}} + 1/2 + \mathcal{O}(e^{-K}),\nonumber\\
&\approx & e^{K_{k}}+1/2.
\end{eqnarray}
This is pretty accurate for any $k$ as long as the turning points of the $k$ th barrier are not too closely spaced. Now with the aid of II-V, we may write  $X^{m_e}_{n}(K)$ for any $m$ as
\begin{eqnarray}
X^{m_c}(K) &\approx &(-1)^m 2^{n'-m'} e^{\sum^{m_c}_{i}2K_{i}}.
\end{eqnarray}
The leading-order term in (\ref{for}) is
\begin{eqnarray}
X^{c}(K) &\approx & 2^{n-m_c} e^{\sum^{n-m_c}_{i}2K_{i}}.
\end{eqnarray}
which yields
\begin{eqnarray}
X^{m}(K) &\approx & (-1)^m 2^{n-m} e^{\sum^{n}_{i}2K_{i}}.
\label{xl}
\end{eqnarray}
Now using I, $(\ref{fe})$ and $(\ref{xl})$ we have
\begin{widetext}
\begin{eqnarray}
f^n(\omega)& \approx & e^{\sum^{n}_{i}2K_{i}} \left ( 2^{n-1} - 2^{n-1}  \sum^{n-1}_{\mathclap{i=1}} \cos{ (2\tilde{L}_i)}
+2^{n-2}\sum_{\mathclap{i_1>i^{s}_2}}\cos{ (2\tilde{L}_{i_1} + 2\tilde{L}_{i^{s}_2})}-
2^{n-3}\sum_{\mathclap{i_1>i^{s}_2>i^{s}_3}} \cos{ (2\tilde{L}_{i_1} + 2\tilde{L}_{i^{s}_2} + 2\tilde{L}_{i^{s}_3})}....\right.\nonumber\\
&& \left. ....+  2^{n-m} (-1)^{m}\sum_{\mathclap{i_1>..>i^{s}_{m}}}\cos{ (2\tilde{L}_{i_1}+...+ 2\tilde{L}_{i^{s}_{m}})}.... + 2 (-1)^{n-1} \sum_{\mathclap{i_1>..>i^{s}_{n-1}}}\cos{ (2\tilde{L}_{i_1}+...+ 2\tilde{L}_{i^{s}_{n-1}})}\right).
\label{cse}
\end{eqnarray}
Using trigonometric identities, a generic sum in (\ref{cse}) can be given as
\begin{eqnarray}
\sum_{\mathclap{i_1>..>i^{s}_{m}}}\cos{ (2\tilde{L}_{i_1}+..+ 2\tilde{L}_{i^{s}_{m}})}=2^{m} \sum_{\mathclap{i_1>..>i_{m}}}\cos{2\tilde{L}_{i_1}}\cos{2\tilde{L}_{i_2}}...\cos{ 2\tilde{L}_{i_m}}.
\end{eqnarray}
This helps us to cast $f(\omega)$ into form
\begin{eqnarray}
f^n(\omega)& \approx & 2^{n-1}  e^{\sum^{n}_{i}2K_{i}} \left ( 1 -  \sum^{n-1}_{\mathclap{i=1}} \cos{ 2\tilde{L}_i}
+ \sum_{\mathclap{i_1>i_2}}\cos{2\tilde{L}_{i_1}}\cos{2\tilde{L}_{i_2}} -
\sum_{\mathclap{i_1>i_2>i_3}}\cos{ 2\tilde{L}_{i_1}}\cos{ 2\tilde{L}_{i_2}}\cos{ 2\tilde{L}_{i_3}}....\right.\nonumber\\
&& \left. ....+   (-1)^{m}\sum_{\mathclap{i_1>..>i_{m}}}\cos{2\tilde{L}_{i_1}}\cos{ 2\tilde{L}_{i_2}}..\cos{ 2\tilde{L}_{i_m}}.... +  (-1)^{n-1} \cos{ 2\tilde{L}_{i_1}}\cos{ 2\tilde{L}_{i_2}}..\cos{ 2\tilde{L}_{i_{n-1}}} \right).
\label{cse2}
\end{eqnarray}
To obtain the closed form of the series note that the terms in the parentheses above can be written in the factorized form
\begin{eqnarray}
(1- \cos{2\tilde{L}_{1}}) \left ( 1 -  \sum^{n-2}_{\mathclap{i \neq 1}} \cos{ 2\tilde{L}_i}
+ \sum_{\mathclap{i_1>i_2\neq 1}}\cos{2\tilde{L}_{i_1}}\cos{2\tilde{L}_{i_2}} -
\sum_{\mathclap{i_1>i_2>i_3 \neq 1}}\cos{ 2\tilde{L}_{i_1}}\cos{ 2\tilde{L}_{i_2}}\cos{ 2\tilde{L}_{i_3}}....\right.\nonumber\\
 \left. ....+   (-1)^{m}\sum_{\mathclap{i_1>..>i_{m}\neq 1}}\cos{2\tilde{L}_{i_1}}\cos{ 2\tilde{L}_{i_2}}..\cos{ 2\tilde{L}_{i_m}}.... +  (-1)^{n-2} \sum_{\mathclap{i_1>..>i_{n-2}\neq 1}} \cos{ 2\tilde{L}_{i_1}}\cos{ 2\tilde{L}_{i_2}}..\cos{ 2\tilde{L}_{i_{n-2}}} \right)
\label{cse4}
\end{eqnarray}
\end{widetext}
Performing this factorization successively for $\tilde{L}_{2}...\tilde{L}_{n-1}$ ultimately yields
\begin{eqnarray}
f^n(\omega)& \approx & 2^{n-1} e^{\sum^{n}_{i}2K_{i}}\prod^{n-1}_{i}(1- \cos{2\tilde{L}_{i}})\nonumber\\
&=& 4 ^{n-1}e^{\sum^{n}_{i}2K_{i}}\prod^{n-1}_{i=1} \sin^2 {\tilde{L}_{i}}
\label{cse4}
\end{eqnarray}
A comparision between (\ref{cse4}) and the full formula for $f(\omega)$ for the external field (\ref{3well}) shows that neglecting lower-order exponentials shifts resonance positions by a tiny amount and, introduces a small shrinking in the resonance widths. Nevertheless (\ref{cse4}) is quite useful for practical purposes since it may accurately yield the tunneling probability  if there are no neighboring transmission resonances in the domain of interest.

\subsection{The case of symmetric potentials}

In the special case where $V(z)$ is composed of identical barriers and identical wells, the semiclassical formula for $f(\omega)$ takes a simple form. For instance, for $n=2$ it is instructive to write (\ref{double}) in the polynomial form:
\begin{eqnarray}
f^2(\omega)&=& 2x (1-y) + 2 x^2 (1-y), \nonumber\\
x&=&e^{2K} , \quad   y=\cos{2\tilde{L}}
\end{eqnarray}
In the following we give $f(\omega)$ for up to $n=6$:
\begin{widetext}
\begin{eqnarray}
f^3(\omega)&=& 2^2\, x^3 (1-y)^2 - 2^2\, x^2 (1-y)(2y-1) + x \, (2y-1)^2, \nonumber\\
f^4(\omega)&=& 2^3 x^4 (1-y) ^3 - 2^3 x^3 (1-y)^2(3y-1) + 2^3 x^2 (1-y)(3y-2)y + 2^3 x\, (1-y)y^2 , \nonumber\\
f^5(\omega)&=& 2^4 x^5 (1-y)^4 - 2^4 \, x^4 (1-y)^3 (4y-1) +  2^2 \,x^3 (1-y)^2 (24y^2-12y-1) - 2^2 \,x^2 (1-y) (4y-1)(4y^2-2y-1) \nonumber\\ && +  x\, (4y^2-2y-1)^2,
\nonumber\\
f^6(\omega)&=&  2^{5} x^6 (1-y)^5 - 2^{5} x^5\, (1-y)^4 (5y-1) +  2^4 x^4\, (1-y)^3 (20y^2-8y-1) -2^4 x^3\, (1-y)^2 (20 y^3-12y^2-3y+1) \nonumber\\  &&  +2 x^2\, (1-y) (4y^2-1) (20 y^2-16y-1)   + 2  x\, (1-y) (4y^2-1)^2.
\label{six}
\end{eqnarray}
\end{widetext}
The first two terms  above have the form 
\begin{eqnarray}
2^{n-1}x^{n}(1-y)^{n-1} - 2^{n-1} x^{n-1}((n-1)y-1) (1-y)^{n-2}\nonumber
\end{eqnarray}
The question of whether the coefficients of the remaining polynomials in $y$ can be given in terms of $n$ is interesting in its own right. Here, it is worth mentioning that if $n$ is even, $f^{n}(\omega)$ seems to have an overlapping transmission resonance at $y=1$ at every order in $x$. Thus for symmetric potentials with an even number of turning point pairs(or at least up to $n=6$),  broadening in specific resonance widths might be expected as the number of the wells increases.


\begin{thebibliography}{999}

\bibitem{sauter}
F. Sauter, ``On the behavior of an electron in a homogeneous electric field
 in Dirac's relativistic theory'',
 Zeit. f. Phys. {\bf 69}, 742 (1931).

\bibitem{he}
W. Heisenberg and H. Euler,
``Consequences of Dirac's Theory of Positrons'',
Z. Phys. {\bf 98}, 714 (1936).

\bibitem{schw}
J.~Schwinger,
``On gauge invariance and vacuum polarization'',
Phys. Rev. {\bf 82} (1951) 664.





\bibitem{eli} The Extreme Light Infrastructure (ELI) project: http://www.extreme-light-infrastructure.eu/eli-home.php



\bibitem{nazo1}
  S.~S.~Bulanov, N.~B.~Narozhny, V.~D.~Mur {\it et al.},
  ``On e+e- pair production by a focused laser pulse in vacuum,''
  Phys.\ Lett.\  {\bf A330}, 1-6 (2004);
   S.~S.~Bulanov, V.~D.~Mur, N.~B.~Narozhny, J.~Nees and V.~S.~Popov,
  ``Multiple colliding electromagnetic pulses: a way to lower the threshold of
  e+e- pair production from vacuum,''
  Phys.\ Rev.\ Lett.\  {\bf 104}, 220404 (2010)

\bibitem{dunne1}
  G.~V.~Dunne, H.~Gies and R.~Sch\"utzhold,
  ``Catalysis of Schwinger Vacuum Pair Production,''
  Phys.\ Rev.\  D {\bf 80}, 111301 (2009)


\bibitem{dunne2}
  R.~Sch\"utzhold, H.~Gies and G.~Dunne,
  ``Dynamically assisted Schwinger mechanism,''
  Phys.\ Rev.\ Lett.\  {\bf 101}, 130404 (2008);



\bibitem{gies1}
F.~Hebenstreit, R.~Alkofer, G.~V.~Dunne and H.~Gies,
  ``Momentum signatures for Schwinger pair production in short laser pulses
  with subcycle structure,''
  Phys.\ Rev.\ Lett.\  {\bf 102}, 150404 (2009)


\bibitem{pizza1}
  A.~Di Piazza, E.~Lotstedt, A.~I.~Milstein and C.~H.~Keitel,
  ``Barrier control in tunneling $e^+ e^-$ photoproduction,''
  Phys.\ Rev.\ Lett.\  {\bf 103}, 170403 (2009)


\bibitem{bellnazo}
  A.~R.~Bell, J.~G.~Kirk,
  ``Possibility of Prolific Pair Production with High-Power Lasers,''
  Phys.\ Rev.\ Lett.\  {\bf 101}, 200403 (2008);
  N.~V.~Elkina, A.~M.~Fedotov, I.~Y.~.Kostyukov, M.V. Legkov, N.B. Narozhny, E.N. Nerush, and H. Ruhl,
  ``QED cascades induced by circularly polarized laser fields,''



\bibitem{anton}
  T.~Heinzl, A.~Ilderton, M.~Marklund,
  ``Finite size effects in stimulated laser pair production,''
  Phys.\ Lett.\  {\bf B 692}, 250-256 (2010).

\bibitem{flor1}
  M.~Orthaber, F.~Hebenstreit and R.~Alkofer,
  ``Momentum Spectra for Dynamically Assisted Schwinger Pair Production,''
Phys.\ Lett.\  B \ {\bf 698}, 80-85 (2011)

\bibitem{dumlu3}
  C.~K.~Dumlu,
  ``Schwinger vacuum pair production in chirped laser pulses,''
  Phys.\ Rev.\  D {\bf 82}, 045007 (2010)


\bibitem{niki1}
N.~ B.~ Narozhny and A.~ I.~ Nikishov,
``Simplest processes in the pair creating electric field,''
Yad.\ Fiz. \ {\bf 11}, 1072 (1970), [Sov. J. Nucl. Phys. 11, {\bf 67} 1503 (1970)]

\bibitem{niki2}
  A.~I.~ Nikishov,
  ``Barrier scattering in field theory removal of Klein paradox,''
  Nuc.\ Phys.\  {\bf 21}, 346 (1970);
 A.~I.~ Nikishov,
  ``Scattering and pair production by a potential barrier,''
  Phys.\ Atom.\ Nucl.  {\bf 67}, 1478 (2004)

\bibitem{hansen}
A.~Hansen, F.~Ravndal
  ``Klein paradox and its resolution ,''
  Phys.\ Script.\ D \ {\bf 23}, 1036-1042 (1981)

\bibitem{gies2}
H.~Gies, K.~Klingmuller
  ``Pair production in inhomogeneous fields ,''
  Phys.\ Rev.\ D \ {\bf 72}, 065001 (2005)


\bibitem{kimpage}
 S.~P.~Kim and D.~Page,
  ``Schwinger pair production via instantons in a strong electric field,''
  Phys.\ Rev.\ D {\bf 65}, 105002 (2002);
  ``Schwinger pair production in electric and magnetic fields,''
 Phys.\ Rev.\  D {\bf 73}, 065020 (2006);
  ``Improved approximations for fermion pair production in inhomogeneous electric fields,''
  Phys.\ Rev.\  D {\bf 75}, 045013 (2007)


\bibitem{remo}
 H.~Kleinert,  R.~Ruffini, S.~S.~Xue,
  ``Electron-Positron Pair Production in Space- or Time-Dependent Electric Fields''
  Phys.\ Rev.\  D {\bf 78}, 025011 (2008)

\bibitem{greinerb}

W.~Greiner, B.~M\"{u}ller, and J.~Rafelski,
{\it , Quantum Electrodynamics of Strong Fields
}(Springer Verlag, Berlin, 1985).


\bibitem{dombey}
A.~ Calogeracos and N.~ Dombey
  ``History and Physics of the Klein Paradox,''
  Contemp.\ Phys.\  {\bf 40}, 313-321 (1999).


\bibitem{mois}
  N.~Moiseyev
  ``Quantum theory of resonances:
calculating energies, widths and cross-sections by complex scaling,''
  Phys.\ Rept.\  {\bf 302}, 211-293 (1998).


\bibitem{sakaki}
  M.~Tsuchiya, T.~Matsusue, H.~Sakaki,
  ``Tunneling Escape Rate of Electrons From Quantum Well in Double-Barrier Heterostructures,''
  Phys.\ Rev.\ Lett. \  {\bf  59}, 2356 (1987).


\bibitem{grob}
  Q.~Z.~Lv. et al,
  ``Noncompeting Channel Approach to Pair Creation in Supercritical Fields,''
  Phys.\ Rev.\ Lett. \  {\bf  111}, 183204 (2013).

\bibitem{meyer}
 B.~Meyer, D.~Vanderbilt
  ``Ab \textit{initio} study of BaTiO$_3$ and PbTiO$_3$ surfaces in external electric fields,''
  Phys.\ Rev.\  B \ {\bf 63}, 205426 (2001).

\bibitem{schg}
D.~Allor, Thomas~D.~Cohen and  David~A.~McGady,
  ``Schwinger mechanism and graphene,''
Phys.\ Rev.\  D {\bf 78}, 096009 (2008)


\bibitem{gra1}
 M.~Barbier,  F.~M.~Peeters,  P.~Vasilopoulos,  J.~M.~Pereira
  ``Dirac and Klein-Gordon particles in one-dimensional periodic potentials''
  Phys.\ Rev.\  B {\bf 77}, 115446 (2008)

\bibitem{gra2}
 C.~Bai,  and X.~Zhang,
  ``Klein paradox and resonant tunneling in a graphene superlattice''
  Phys.\ Rev.\  B {\bf 76}, 075430 (2007)

\bibitem{gra3}
 J.~M.~Pereira,   P.~Vasilopoulos,  F.~M.~Peeters,
  ``Graphene-based resonant-tunneling structures''
  Appl.\ Phys.\  Lett. {\bf 90}, 132122 (2007)





New York, 1995).





\bibitem{popov1}
V.~S.~Popov,
``Pair Production in a Variable External Field (Quasiclassical approximation)'',
Sov. Phys. JETP {\bf 34}, 709 (1972);
``Pair production in a variable and homogeneous electric field as an oscillator problem''.
Sov. Phys. JETP {\bf 35}, 659 (1972)



\bibitem{motto}
  Y.~Kluger, J.~Eisenberg, B.~Svetitsky, F.~Cooper and E.~Mottola,
  ``Pair production in a strong electric field,''
  Phys.\ Rev.\ Lett.\  {\bf 67}, 2427 (1991);

   ``Fermion Pair Production In A Strong Electric Field,''
  Phys.\ Rev.\  D {\bf 45}, 4659 (1992);

Y.~Kluger, E.~Mottola and  J.~Eisenberg,
  ``The quantum Vlasov equation and its Markov limit,''
Phys.\ Rev.\  D {\bf 58}, 125015 (1998)


\bibitem{heading}

J.~Heading, {\it An Introduction to Phase-Integral Methods}, (Methuen, London, 1962).


\bibitem{froman1}

N.~Froman and P.~O.~Froman, {\it Physical Problems Solved by the Phase-Integral Methods},  (Cambridge University Press,
Cambridge, 2004).




\bibitem{wood}
 P.~Kennedy
  ``The Woods-Saxon potential in the Dirac equation''
  J.\ Phys.\  A \ Math. \ Gen. {\bf 35}, 689-698 (2002)





\bibitem{froman2}

N.~Froman and P.~O.~Froman, {\it Phase Integral Method Allowing Nearlying Transition Points},  (Springer-Verlag, New York
 1998).



\bibitem{bender}
 C.~M.~Bender, K.~Olaussen, P.~S.~Wang
  ``Numerological analysis of the WKB approximation in large order,''
  Phys.\ Rev.\  D \ {\bf 16}, 1740 (1977).



\bibitem{dumlu2}
  C.~K.~Dumlu and G.~V.~Dunne,
  ``Stokes phenomenon and Schwinger vacuum pair production in time dependent laser pulses,''
  Phys.\ Rev.\ Lett.\ {\bf 104}, 250402 (2010)

\end{thebibliography}
\end{document}